\documentclass[prd,showpacs,showkeys,nofootinbib,floatfix,
               preprint,12pt,tightenlines,fleqn]{revtex4}

\usepackage{amsmath,amsfonts,latexsym,graphicx,mathrsfs}

\newcommand {\version}{v5}

\newcommand{\st} {spacetime}
\newcommand{\stf}{spacetime foam}
\newcommand{\dr} {dispersion relation}

\newcommand{\lP} {l_\text{Planck}}

\newcommand{\beq}{\begin{equation}}
\newcommand{\eeq}{\end{equation}}
\newcommand{\beqa}{\begin{eqnarray}}
\newcommand{\eeqa}{\end{eqnarray}}

\def\d{\mathrm{d}}
\newcommand{\half}{{\textstyle \frac{1}{2}}}

\newcommand{\rhs} {right-hand side}

\newcommand{\YMth} {Yang--Mills theory}

\begin{document}
\noindent JETP Lett. 86, 73 (2007)\hfill
gr-qc/0703009 (\version)\newline\vspace*{1\baselineskip}
\title{Fundamental length scale of quantum spacetime foam}
\author{F.R.\ Klinkhamer}
\email{frans.klinkhamer@physik.uni-karlsruhe.de}
\affiliation{Institute for Theoretical Physics, University of Karlsruhe (TH),
            76128 Karlsruhe, Germany
            \\
            }

\begin{abstract}
\vspace*{.25\baselineskip}\noindent
It is argued that the fundamental length scale for the
quantum dynamics of spacetime need not be equal to the Planck length.
Possibly, this new length scale is related to
a nonvanishing cosmological constant or vacuum energy density.
\end{abstract}



\maketitle

\section{Introduction}
\label{sec:intro}

It is generally assumed that the structure of a quantum
\stf~\cite{Wheeler1957,Wheeler1968,Hawking1978,Hawking1979},
if physically relevant, is given by a \emph{single} fundamental
length scale, the Planck length $\lP \equiv\sqrt{\hbar\,G/c^3}$.
In this Letter,
we point out that a pure quantum \stf~(``pure'' meaning independent
of the direct presence of matter or nongravitational fields) could
have a fundamental length scale $l$ different from
$\lP \propto \sqrt{G}$. Heuristically,
the quantum \stf~would arise from gravitational self-interactions
which need not involve Newton's constant
$G$ describing the gravitational coupling of matter
(similar to the case of a gas of instantons in \YMth~\cite{Eguchi1980},
where the existence of gauge-field instantons does not depend on
the coupling to additional fermion fields).

Henceforth, we adopt the conventions of Ref.~\cite{Weinberg1972},
set the velocity of light in vacuum $c$  to unity,
and consider \st~manifolds without boundaries (so that boundary terms
in the action need not be considered explicitly \cite{Hawking1979}).
The ``matter'' Lagrangian density is denoted by $\mathcal{L}_\text{M}(x)$
and we follow Einstein \cite{Einstein1917} by
introducing a nonzero (positive) cosmological constant $\lambda$
with the dimensions of an inverse length square.
The standard classical action of general relativity (GR)
is then given by  \cite{Weinberg1972}:
\beqa\label{eq:Sstandard}
S_\text{\,gravitation}^\text{\,standard} &=&
- \frac{1}{16 \pi \,G} \int \d^4 x \,
\sqrt{|g(x)|}\; \big( R(x)+2\lambda \big)
+ \int \d^4 x \, \sqrt{|g(x)|}\; \mathcal{L}_\text{M}(x)\;,
\eeqa
in terms of the Ricci scalar $R(x) \equiv g_{\mu\nu}(x)\,R^{\mu\nu}(x)$
and the scalar density $g(x) \equiv \det g_{\mu\nu}(x)$.
Varying the metric $g_{\mu\nu}(x)$, the stationary-action principle
gives the Einstein field equations \cite{Weinberg1972,Einstein1917}:
\beq\label{eq:classical-field-equations}
R^{\mu\nu}(x) - \half\, g^{\mu\nu}(x)\, R(x) - \lambda\,g^{\mu\nu}(x)
= - 8\pi \, G \, T^{\mu\nu}(x)\;,
\eeq
where the energy-momentum tensor $T^{\mu\nu}(x)$ of the matter
is defined by the following functional derivative:
\beq\label{eq:Tmunu}
T^{\mu\nu}(x) \equiv \frac{2}{\sqrt{|g(x)|}}\;\;
\frac{\delta\left(\textstyle{\int}\d^4 y\,\sqrt{|g(y)|}\;
\mathcal{L}_\text{M}(y)\right)}
     {\delta g_{\mu\nu}(x)}\,.
\eeq

The experimental tests of GR \cite{Weinberg1972,Will1993} only refer to
solutions of the classical field equations
\eqref{eq:classical-field-equations}.
The reason is that classical gravity is already extraordinarily weak, so
that even smaller quantum corrections are totally out of reach experimentally.
Still, there are certain theoretical investigations
which go beyond the classical theory; however, they
essentially correspond to
a semiclassical version of \eqref{eq:classical-field-equations},
with $T^{\mu\nu}$ on the \rhs~replaced
by a (renormalized) vacuum expectation value of the appropriate
expression in terms of quantum fields defined over a classical
\st~background \cite{Birrell1982}.
Thus, as far as gravity is concerned,
we can only be sure of the classical field equations
\eqref{eq:classical-field-equations}.

The standard form of the action has the following defining property.
Setting the cosmological constant to zero, $\lambda=0$,
the action \eqref{eq:Sstandard}
for Minkowski spacetime $g_{\mu\nu}(x)=\text{diag}(-1,1,1,1)$
reduces precisely to the special-relativity action of the matter fields
(in manifest agreement with the Equivalence Principle \cite{Veltman1976}).
But, in this Letter, we are \emph{only} interested in gravitational
effects, taking the classical matter fields for granted.
Moreover, we intend to stay completely agnostic
as regards the quantum effects of spacetime and take
the cosmological constant $\lambda$, viewed as a geometrical effect,
to be strictly nonzero.

\section{Generalized action}
\label{sec:generalized-action}

Even though the theory of ``quantum gravity'' does not exist, we can try
to make some general remarks starting from the rescaled action which
enters the quantum world of probability amplitudes via the Feynman
\cite{Feynman1948} phase factor $\exp(\mathrm{i}\,\mathcal{I})$,
with $\mathcal{I}=S/\hbar$ expressed in terms of the classical action $S$
and the reduced Planck constant $\hbar \equiv h/2\pi$.
This phase factor would, for example, appear in a properly
defined path integral; see Ref.~\cite{Hawking1979} for a discussion
in the Euclidean framework.

The fact remains, however, that our understanding of the merging of
quantum mechanics and gravitation
is far from complete; see Ref.~\cite{Penrose1996}
for a clear account of at least one open problem.
It may, therefore, be sensible to consider a \emph{generalized}
dimensionless action as possibly being relevant to a
future quantum-gravity theory:
\beqa\label{eq:Sgeneralized}
\mathcal{I}_\text{\,gravitation}^\text{\,generalized} &=&
- \frac{1}{16 \pi \,l^2} \int \d^4 x \,
\sqrt{|g(x)|}\; \big( R(x)+2\lambda \big)
+ \frac{G/c^3}{l^2} \int \d^4 x \,
\sqrt{|g(x)|}\; \mathcal{L}_\text{M}(x)\;,
\eeqa
where $l$ is a new fundamental length scale and $c$ has been temporarily
restored.
To emphasize once more, the generalized action \eqref{eq:Sgeneralized}
is only to be used for a rough description of possible quantum effects
of \st, not those of the matter fields. Specifically, the matter fields
which enter $\mathcal{L}_\text{M}(x)$ in \eqref{eq:Sgeneralized}
are considered to act as fixed classical sources and cannot be rescaled.
In fact, $\hbar$ does not appear at all in \eqref{eq:Sgeneralized},
as will be discussed further below.

Setting $l$  equal to
the Planck length obtained from Newton's gravitational constant $G$, the
light velocity $c$, and Planck's quantum of action $h\equiv 2\pi\,\hbar$,
\beq\label{eq:lPlanck}
\lP \equiv \sqrt{\hbar\,G/c^3} \approx 1.6 \times 10^{-35}\,\text{m}\;,
\eeq
the standard form \eqref{eq:Sstandard} of the action
is reproduced (taking again $c=1$):
\beq\label{eq:S-for-l-equal-lPlanck}
\mathcal{I}_\text{\,gravitation}^\text{\,generalized}\,\Big|_{l=\lP}=
S_\text{\,gravitation}^\text{\,standard} /\hbar\;.
\eeq
Yet, the generalized action \eqref{eq:Sgeneralized}
with an independent length scale $l$
looks more natural, as Newton's constant $G$ multiplies the classical
source term. The overall factor $l^{-2}$ in \eqref{eq:Sgeneralized}
is, of course, irrelevant for obtaining the classical field
equations \eqref{eq:classical-field-equations}.
But, as stressed by Feynman \cite{Feynman1948}, quantum physics is
governed by the phase factor $\exp(\mathrm{i}\,\mathcal{I})$ and,
for a genuine quantum spacetime,
the overall factor of $l^{-2}$ in \eqref{eq:Sgeneralized}
would be physically relevant.

The expression \eqref{eq:Sgeneralized} for the gravitational
quantum phase also suggests that, as far as \st~is concerned,
the role of Planck's constant $\hbar$ would be replaced by
the squared length $l^2$, which might loosely be called
the ``quantum of area.''
Planck's constant $\hbar$ would continue to play a role in
the description of the matter quantum fields.
But, with $\hbar$ and $l^2$ \emph{logically independent}, it is possible to
consider the ``limit'' $\hbar \to 0$ (matter behaving classically) while
keeping $l^2$ fixed (spacetime behaving nonclassically),
which would correspond to the applicability domain
of the generalized action \eqref{eq:Sgeneralized}.
At the level of \emph{Gedankenexperiments},
the quantum spacetime resulting from \eqref{eq:Sgeneralized} could
then be studied with classical measuring rods and standard clocks.

Table~\ref{tab:table1} summarizes the (overcomplete) set
of fundamental dimensionful constants at our disposal.
The scope of this Letter has deliberately been restricted to a
qualitative discussion of the second and third columns of
Table~\ref{tab:table1}, thus, leaving the unified and rigorous
treatment of \emph{all} columns to a future theory.
For example, the Hawking temperature of a Schwarzschild black hole
with mass $M$ is given by $T_\text{\,H}=\hbar\, c^3/(8\pi k G M)$
and lies outside the scope of the present Letter ($\hbar=0$);
however, a future quantum-gravity theory
(possibly with fundamental constants $\hbar$, $c$, $G$, and $l^2\,$)
would certainly have to explain black-hole thermodynamics \cite{Wald1995}.

Referring to Table~\ref{tab:table1}, two parenthetical remarks
can be made, one on the quantum of area entering the third column and
another on a possible type of theory for all of the columns of the table.
First, observe that  loop-quantum-gravity
calculations \cite{Rovelli1995,Ashtekar2004},
with $\lP$ replaced by $l$, give for a subset of the eigenvalue spectrum
of the area operator the values
$8\pi\gamma\,l^2\sum_{i=1}^{n} \sqrt{j_i\,(j_i+1)}$, with positive
half-integers $j_i$, positive integers $n$, and a real parameter $\gamma>0$.
Very briefly, $j_i$ labels the irreducible $SU(2)$ representation of
spin-network link $i$ intersecting the test surface $\mathcal{S}$ and the sum
over $i$ builds up the area of $\mathcal{S}$, with the (somewhat mysterious)
Barbero--Immirzi parameter $\gamma$ entering the definition of the
quantum theory. The smallest eigenvalue in this subset is apparently
given by $4\pi\sqrt{3}\,\gamma\,l^2$, which would then correspond to the
precise value of the quantum of area.
The fundamental role of a quantum of \emph{area}
would be in-line  with the suggested ``holographic principle''
\cite{'tHooft1993,Susskind1994}.

\begin{table}[t]
\caption{\label{tab:table1} Fundamental constants of nature,
including the hypothetical quantum of area $l^2$.}
\begin{ruledtabular}
\begin{tabular}{ccc}
\hspace*{10mm}
\hspace*{5mm}quantum
\hspace*{5mm}
\hspace*{10mm}
&
\hspace*{5mm}classical
\hspace*{5mm}
\hspace*{10mm}
&
quantum
\hspace*{5mm}
\hspace*{10mm}
\\[-1mm]
\hspace*{10mm}
\hspace*{5mm}matter
\hspace*{5mm}
\hspace*{10mm}
&
\hspace*{5mm}relativity
\hspace*{5mm}
\hspace*{10mm}
&
spacetime
\hspace*{5mm}
\hspace*{10mm}
\\
\hline
\hspace*{10mm}
\hspace*{5mm}$\hbar$
\hspace*{5mm}
\hspace*{10mm}
&
\hspace*{5mm}$c\,,\,G$
\hspace*{5mm}
\hspace*{10mm}
&
\hspace*{2mm}$l^2$
\hspace*{5mm}
\hspace*{10mm}
\\
\end{tabular}
\end{ruledtabular}
\end{table}

Second, observe that,  with an extra fundamental length $l$ available
in Table~\ref{tab:table1}, it would be possible to eliminate, for example,
Newton's gravitational constant $G$
by writing $G=f\,c^3\,l^2/\hbar\,$ with a numerical factor $f$.
This trivial exercise may hint at a new type of induced-gravity theory
\cite{Sakharov1967,Visser2002}, with ``classical gravitation'' emerging
from the \emph{combined} quantum effects
of matter and spacetime, so that the number $f$ may be calculable.
In fact, the magnitude of the Newtonian gravitational acceleration
towards a macroscopic point mass $M$ at a macroscopic
distance $D$ can be written in the following suggestive form:
$G M/D^2 = f\,c\, (Mc^2/\hbar)\, (l^2/D^2)$,
with all microscopic quantities indicated by lower-case symbols.
Remarkably, the possible appearance of a new parameter in
quantum-gravity theory has also been suggested by a
hydrodynamics analogy \cite{Volovik2006}.

Returning to the earlier discussion of $l$ and $\lP$,
the only conclusion, for the moment, is that it may be important to
recognize the length scale $l$ as a free parameter of the quantum theory
of spacetime, which is not directly related to $\lP$.
Putting aside epistemological issues, there are then two cases to discuss:
$l$ larger than or equal to $\lP$ and $l$ smaller than $\lP$.

The first case ($l \geq \lP$) appears to be a serious physical possibility.
Presupposing the existence of a classical \st~manifold for reference,
$\lP$~would correspond to the minimal length which could, in principle,
be measured by a
massive particle obeying the Heisenberg position--momentum uncertainty
relations; cf. Refs.~\cite{Mead1964,Garay1994}.
But, as the probed distances are made shorter and shorter
(starting at the atomic scale, for example), it could be that
the classical \st~picture breaks down \emph{well before} the length $\lP$
is reached, the breakdown occurring at distances of order $l$
(assuming, for the moment, that $l \gg \lP$).

The second case ($l < \lP$) looks perhaps rather academic
as long as all the material probes of the spacetime
interact with a coupling strength
$G$. However, it could be (possibly in an extended version of the theory)
that a sub-Planckian \st~structure determines certain effective
parameters for the physics over distances of the order of $\lP$~or
larger. An analogy would be atomic physics, which  determines
the electric permittivity $\epsilon$ and magnetic permeability $\mu$
controlling  the propagation
of electromagnetic waves with wavelengths very much larger than the
atomic length scale.

In either case, the main outstanding problem is to understand
how the \emph{quantum} \stf~gives rise to an effective
\emph{classical} spacetime manifold over large enough distances,
perhaps requiring an objective state-reduction mechanism
(cf. Refs.~\cite{Penrose1996,Penrose2005}).
But, given our daily experience, we can be sure that  there must be
some kind of ``crystallization'' of classical \st, with or without
the occasional ``defect'' in the resulting structure.

Incidentally,  a nontrivial classical spacetime-foam remnant
can be expected to lead to an effective
violation of the Weak Equivalence Principle (WEP) \cite{Will1993}
on the following grounds. Lorentz invariance of a massless spin--2
particle (graviton) essentially implies WEP \cite{Veltman1976,Weinberg1964}.
But soft gravitons propagating over a classical spacetime-foam remnant
are believed to have Lorentz violation
(for example, from a modified dispersion relation)
and, therefore, WEP is no longer guaranteed to hold.
Moreover, particles of different spin can be expected to propagate
differently over a classical \st~with nontrivial small-scale structure,
because of different  ``boundary conditions'' on the particle fields.
Precisely these type of propagation effects can lead to interesting
astrophysical bounds, as will be briefly discussed in the next section.

\section{Conjecture}
\label{sec:conjecture}

The crucial question, now, is what the value of the
fundamental length scale $l$ really is, compared to $\lP$?
At this moment, the answer is entirely open and
$l = \lP$ is certainly not excluded.
Let us, here, present a conjecture suggested by cosmology. Henceforth,
we set $c=\hbar=k=1$, where $k$ is the Boltzmann constant.

Recent astronomical observations seem to indicate a nonzero value
of the cosmological constant, $\lambda_0 \equiv L_0^{-2} > 0$, with
a length scale $L_0$ corresponding to the size of the visible universe,
$L_0 \approx 10^{10}\,\text{ly}
\approx 10^{26}\,\text{m}$;
see Refs.~\cite{Spergel2006,Riess2006} and references therein.
In addition, it is possible that the universe at an early
stage had a significant vacuum energy density,
$\rho_\text{vac} \equiv E_\text{vac}^4$;
see, e.g., Ref.~\cite{Mukhanov2005} for
background on so-called ``inflation'' models.
Setting $\lambda$ in \eqref{eq:Sgeneralized} equal to $\lambda_0$
and $\mathcal{L}_\text{M}(x)$ to $-\rho_\text{vac}$
(that is, neglecting the
kinetic terms of the fields in first approximation), one has
\beqa\label{eq:S-universe}
\mathcal{I}_\text{\,gravitation}^\text{\,generalized} &\sim&
- \frac{1}{16 \pi \,l^2} \int \d^4 x \,
\sqrt{|g(x)|}\;
\big( R(x)+2\lambda_0 +2\Lambda \big) \;,
\eeqa
with the effective cosmological constant
\beq\label{eq:Lambda}
\Lambda \equiv  8 \pi \,\lP^2\,\rho_\text{vac}
= 8 \pi \, E_\text{vac}^4/E_\text{Planck}^2 \;,
\eeq
in term of  the energy scales $E_\text{vac}$ and
$E_\text{Planck} \equiv 1/\lP \approx 1.2 \times 10^{19}\,\text{GeV}$.

For the case $\Lambda \gg \lambda_0$ (possibly relevant
to the early universe), it may be natural to have
the lengths  $l$ and  $\sqrt{3\pi/\Lambda}$ of the same order,
where the numerical factor $\sqrt{3\pi}$ has been introduced to
streamline the discussion.
The de-Sitter instanton \cite{Hawking1979,Eguchi1980}, for example,
gives for the right-hand side of \eqref{eq:S-universe} an absolute value of
$3\pi/(l^2\,\Lambda)\,[1+\text{O}(\lambda_0/\Lambda)]$,
which would then be of order unity. As mentioned in the Introduction,
the quantum \st~foam would somehow resemble a gas of these
gravitational instantons.

For the case $0 \leq \Lambda \leq \lambda_0$
(possibly relevant to the
final state of the universe), the same argument suggest that
the fundamental length scale  $l$ would be
of the order of $\sqrt{3\pi/\lambda_0}$, independent of both $\hbar$ and $G$.
Our galaxy would then be buried \emph{deep inside}
a quantum-spacetime-foam remnant with
typical length scale $l \sim 3\, L_0$ $\approx$ $3 \times 10^{26}\,\text{m}$,
provided the effective classical \st~manifold can be smoothed
over distances less than $l$. The resulting picture of the universe
would, in a way, resemble that of Linde's ``chaotic inflation''
\cite{Mukhanov2005,Linde1983} but have a different origin.

Returning to the case $\Lambda \gg \lambda_0$,
the conjecture is, therefore, that the fundamental length scale $l$
can be significantly larger than the Planck length
and have the following order of magnitude:
\beqa\label{eq:l-conjecture}
\hspace*{-1mm} l &\stackrel{?}{\sim}& E_\text{Planck}/E_\text{vac}^2
              \approx 
\hspace*{-1mm} 1.6 \times 10^{-29}\,\text{m}\,
\left( \frac{E_\text{Planck}}{10^{19}\,\text{GeV}} \right)\,
\left( \frac{10^{16}\,\text{GeV}}{E_\text{vac}} \right)^2\,,
\eeqa
where the numerical value for $E_\text{vac}$
has been identified with the ``grand-unification'' scale
suggested by elementary particle physics \cite{GeorgiQuinnWeinberg1974}.
(The energy scale $E_\text{Planck} \equiv 1/\lP \gg 1/l$
would continue to play a role in, for example,
the ultra-high-energy scattering of massive neutral particles
by graviton exchange.)
In terms of the fundamental constants $G$ and $l$ from the
generalized action \eqref{eq:Sgeneralized}, the vacuum energy density would
be given by (temporarily reinstating $c$)
\beq\label{eq:rho-vac}
\rho_\text{vac} \stackrel{?}{\sim} c^4/\left(G\,l^2 \right)\;,
\eeq
which would suggest a gravitational/\st~origin for this vacuum energy density
($\hbar$ not appearing directly; cf. Table~\ref{tab:table1}).

Remark that the detailed quantum structure of \st~may very well
involve \emph{two} length scales, here taken to be $\lP$ and $l$.
Indeed, if \eqref{eq:l-conjecture} holds true with $\lP / l \sim 10^{-6}$,
it is perhaps possible to have sufficiently rare spacetime-foam defects
left-over from the crystallization process mentioned
in the penultimate paragraph of Sec.~\ref{sec:generalized-action}.
With average \st~defect size $\widetilde{b}$ set by $\lP$
and average separation $\widetilde{l}$ set by $l$,
these time-dependent defects would be close to saturating the current
astrophysical bounds from ultra-high-energy cosmic rays,
the so-called excluded-volume factor
in the modified photon \dr~being
$(\,\widetilde{b}/\widetilde{l}\,)^4 \sim 10^{-24}$;
see Ref.~\cite{BernadotteKlinkhamer2007} for details.
(The present Letter originated, in fact, from an attempt to understand
these astrophysical bounds.)

Equally important, with $l \gg \lP$, the dynamics of the early universe
for temperatures $T \gtrsim 1/l$ $\stackrel{?}{\sim} 10^{13}\,\text{GeV}$
would have to be reconsidered as a classical \st~description might
be no longer relevant. For sufficiently low temperatures
($T \ll 1/l$), where a classical \st~description should be valid,
the expansion of the universe would still be given by the standard 
Einstein equations \eqref{eq:classical-field-equations}
and corresponding Friedmann equations \cite{Weinberg1972},
as the generalized quantum action \eqref{eq:Sgeneralized}
has been designed not to modify the basic classical field equations.

In this section, we have presented only one particular conjecture
for the fundamental length scale $l$. But the idea is more
general, namely that entirely new physics may be responsible for
the small-scale structure of \st.

\noindent\section*{ACKNOWLEDGMENTS}
The author is grateful to C.~L\"{a}mmerzahl for comments on the manuscript
and to C.~Kaufhold for help with obtaining some of the references.


\end{document}